# Novel Technosignatures

**A. A. Jackson**
Triton Systems LLC, Houston, USA,
al_jackson@aajiv.net
**Gregory Benford**
Department of Physics & Astronomy, USA, UC Irvine, xbenford@gmail.com

*My rule is there is nothing so big nor so crazy that one out of a million technological societies may not feel itself driven to do, provided it is physically possible.*
— *Freeman Dyson*

Abstract
Technosignatures can represent a sign of technology that may infer the existence of intelligent life elsewhere in the universe. This had usually meant searches for extraterrestrial intelligence using narrow-band radio signals or pulsed lasers. Back in 1960 Freeman Dyson put forward the idea that advanced civilizations may construct large structures in order to capture, for use, the energy of their local star, leading to an object with an unusual infrared signature. Later it was noted that other objects may represent the signature of very advanced instrumentalities, such as interstellar vehicles, beaming stations for propulsion, unusual beacons not using radio or laser radiation but emission of gamma rays, neutrinos or gravitational radiation. Signs may be unintentional or may be directed. Among directed and undirected signs we present some models for signaling and by-product radiation that might be produced by extremely advanced societies not usually considered in the search for extraterrestrial intelligence.

## I. Introduction

The most studied approach to SETI is by way of the electromagnetic spectrum, mostly radio and possibly lasers, the infrared being favored. Many new methods of doing SETI are in the works [1], but one can ask the question: are there any other signatures of advanced extraterrestrial civilizations?

At almost the time of the paper by Philip Morrison and Giuseppe Cocco [2] Freeman Dyson [3] and Nikolai Kardashev [4] noted that the artifacts of advanced civilizations with innovative technologies could build artifacts such as Dyson Spheres or Kardashev civilizations which may have observable properties.
Briefly the Kardashev classification is:
  Type I – harnesses the energy output of an entire planet.
  Type II – harnesses the energy output of a star, and generate about 10 billion times the energy output of a Type I civilization.
  Type III – harnesses the energy output of a galaxy, or about 10 billion times the energy output of a Type II civilization.
  Roughly then a Dyson Sphere would represent the technology of a type II Kardashev civilization.
  In following Kardashev I, II and III civilizations are denoted as K1, K2 and K3. (Note: Strictly speaking Kardashev's original paper dealt with how an advanced civilization might power interstellar 'beacons'. Informally his classification has passed into a scheme of taxonomy for tagging advanced civilizations, wither that is a correct thing to do will not be debated here.)
  All materials composing a Dyson sphere would radiate waste heat in the infrared (or longer wavelengths) of the electromagnetic spectrum [4a]. Searches have been made for candidate Dyson Sphere but no definitive identification has been made [4a]. Just what kind of a technology one might look for at K3 scales has been quantified, in the case of Galaxy wide Dyson spheres but nothing seen [4a], it is not entirely clear what K3 signatures are worth looking for.
  We explore other exotic possibilities of signatures by advanced civilizations in the following.

## II. Starships

Consider K1 and K2 civilizations building starships. Might these be detectable in parts of the electromagnetic spectrum not usually associated with SETI? Viewing, Horswell and Palmer [5] asked such a question in 1977. They enumerated the possibilities:
1. Innocuous - Slow interstellar flight, such as World Ships.
2. Energetic
    i. Nuclear Fission
    ii. Thermonuclear Fusion
    iii. Matter-Antimatter

Viewing, et al., did not draw any particular conclusions about the quantified detectability.
Zubrin 1995[5a] examined the same question of energetic starships and did put forward some examples of detection. His considerations are given in Table 1:

Table 1 Observable Starships

| Type | Radiated at Source | Frequency | Detection Object |
|---|---|---|---|
| Radio | 80-2000 TW | 24 – 48 kHz | Yes- Magsails |
| Visible | 120000 TW | IR | Yes – Nuclear 300 ly |
| X-Rays | 40000 TW | 2 - 80 KeV | Nuclear and Antimatter- Ships ~10 ly-1000ly |
| Gamma Rays | 1 – 32 MeV | 20-200 Mev | Antimatter Ships |

Assumptions were made about mass and acceleration of the vehicle; consult the paper [5a].
Consider beaming stations which propel sails or similar arrays. Civilizations using beamed radiation, a straight forward and technologically attractive way of implementing interstellar travel.

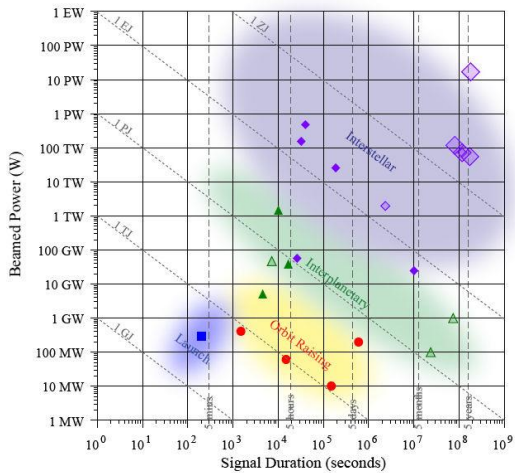

Figure 1: Candidate Beaming Stations (Benford and Benford ( Astrophysical Journal, Volume 825, Number 2, 2016)

In this case we would be looking for the transmitter stations attenuated at whatever distance they are at. Many varieties of radiation may be involved; laser beam power and microwaves have received great attention, Figure 1.

A caveat: in most of the star ships to be the observer has to be inside the transmitter cone of an energy beam. In general this stream of energy will be narrow. If one compares this with the full sky which would be four pi steradians the ratio of beam diameter to the expanse of the sky implies rather small observational probabilities'.

II.1 Relativistic Ships

Following the lead of Freeman Dyson and Nikolai Kardashev we extrapolate.
Take the civilization to be Kardashev 2, or K2, these ships will be taken to be relativistic starships.
1) They can run 'hot' … ship construction materials that can come into thermal equilibrium with temperatures as high as 5000 K (this close to the melting point of grapheme).
2) Material structural strength limits have been overcome, that is there is support Lorentz factors of up to at least 500 or 0.999998 speed of light. This means stress transmitted by drag due to interacting electromagnetic fields or the support of very large magnetic flux densities have been solved.
3) K2 civilizations fly 1g , maybe higher g, ships.
4) Disintegration due to relativistic dust or gas impact or drag in the interstellar medium…. solved.
5) K2 guidance, navigation and control, almost magic but still distinguishable.
6) Whatever the technical problem .. Likely a K2 civilization can solve it.

Postulate a generic K2 ship, a high Lorentz factor ship (that is a large gamma).
Note a Lorentz factor (gamma) of 10 is equivalent to a ship speed of .995 the speed of light.
Take a hypothetical numerical example. Postulate a K2 ship with gamma of 500 (yes that's a 'super science' ship) 0.999998 the speed of light. This hypothetical K2 will be taken to be as hot as 5000 degrees K (carbon like materials have upper limit thermal properties such as this).
Suppose such a star ship is making an interstellar trip, what might we see? While the ships engine is running and even after propulsion is off there will be waste heat. It can be modeled as isotropic radiation in the rest frame of the ship. If $\varepsilon$ is the emissivity (1 for a black body) and σ is the Stefan-Boltzman constant then the energy flux density is $j = \varepsilon \sigma T^4$ (watts/meter$^2$), in the rest frame of the vehicle. If v is the ship velocity and c the speed of light then $\gamma = 1/\sqrt{1-\beta^2}$, where β = v/c, γ is the Lorentz factor.

To an observer in another inertial frame the radiation will be beamed, the relativistic 'headlight' effect, see figure 2.

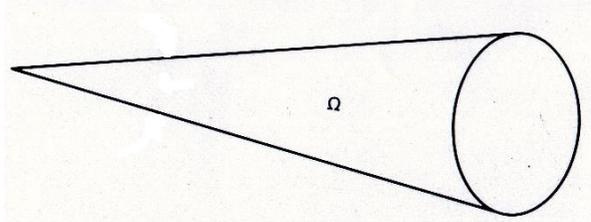

Figure 2. Beamed Radiation for Relativistic Star Ship. Ω = opening solid angle.

The flux density j in the proper frame will be 'Doppler Boosted', to $j_o$ an observer's frame [6]

$$j_o = j\delta^4 \quad \frac{watts}{meter^2} \quad [1]$$

$$\delta = \left(\frac{1+\beta}{1-\beta}\right)^{1/2} . \quad [2]$$

Considering a modest ship of size and mass, a K2 ship accelerating at one gravity up to a γ = 500. For instance a ship 1000 meters long and 50 meters in diameter radiating black body waste heat will be generating 11402 terra-watts in its rest frame, Doppler boosting[7] will generate = $2 \times 10^{16}$ terra watts beamed into the forward direction! However unless the ship is headed straight at the observer it will be almost impossible to see.

The opening solid angle is $\Omega \sim \frac{1}{\gamma}$ (steradians) thus the probability of observation is Ω/4π or about .002. The probability of observation will be difficult.

This is example is very extreme, comparable to x-ray buster EXO 0531-66[7a]. The effect is interesting, consider that 1 watt of black body radiation in a ship's rest frame is Doppler boosted by the observer's frame by $\gamma^4$ this would be a large flux in the frame of an observer.

For this case if one takes into account the Doppler shifting of the characteristic wavelength, from near green in the rest frame or the ship to soft x-rays in the observers frame one may have to rely of satellite observatories in Earth orbit.

Thus one might look for small anomalies in the host of new astrophysical satellite observatories, see list in figure 3.

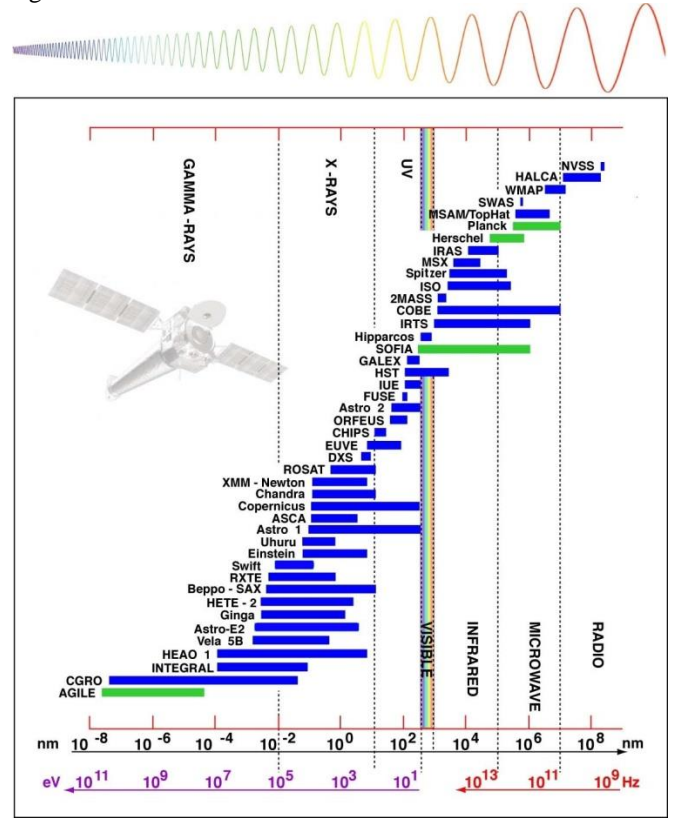

Figure 3 : Astrophysical Observatories and the Electromagnetic Spectrum (source Wikipedia)

III. Gravitational Machines

In 1963 Freeman Dyson [8] suggested that an advanced civilization might use massive binaries as 'slingshots'. A process used by spacecraft in the solar system, in astrodynamics called a Gravity Assist to save fuel and time. Dyson considered white dwarf binaries and neutron star binaries. To these one can add black hole binaries.

Like Dyson take the orbital distance the objects to be circular with a semi major axis of 1000 km.
Consider a ship approaching with a velocity V. Velocity gains then are of the order of .002 to .006 c. Not bad for free energy, except one has to live in the vicinity of or travel to such objects.
There is, however, bad news. The lifetimes, t, of these binaries against gravitational wave energy loss and hence orbit decay to collapse is given by [9]:

$$t = \frac{5c^5 r^4}{512 G^3 m^3} \quad [3]$$

If both binaries have the same mass, m, where c is the speed of light, G the gravitational constant and r the distance between the binaries then for the separation r = 1000 km the lifetimes are

> White Dwarfs ~ 30 years
> Neutron Stars ~ 18 years
> Black Holes ~ .1 year

Larger orbital distances have larger lifetimes but much smaller velocity gains. Achieving high fractions of the speed of light does not look promising for Dyson gravitational machines.

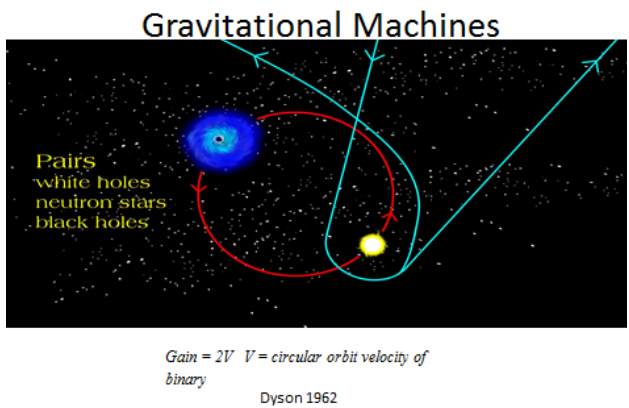

Figure 4 Gravitational Machines (Douglas Potter)

II.I <u>Surfing Black Holes</u>

Another place to look is isolated black holes. Rotating black holes (these will be referred to as Kerr black holes) and non-Rotating black holes have an interesting property when a particle has a trajectory close to black holes, it no longer moves according to Newtonian mechanics.

In Newtonian physics when a spacecraft approaches a planet with a speed at infinity that exceeds escape velocity, from that planet, unless that craft fires a rocket motor, encounters a planet's atmosphere, hits the surface, or uses some other dissipative mechanism it will return to infinity (for example, a parabolic or hyperbolic orbit). However in the case of a black hole when one gets close enough there are orbits that can go into temporary capture. If the Schwarzschild radius is, $r_s = \frac{2GM}{c^2}$, then if a particle's encounter distance is less than $10 r_s$ the motion is strongly non-Newtonian [10].

This article will only be concerned with trajectories (or more correctly time-like and null geodesics (photons)) that are initially unbound, that is that come in from infinity and have an impact parameter b. In Newtonian mechanics a particle has a total energy E then particles with E > 0 will be remain on unbound orbits (if they don't hit their central gravitating body) and with E < 0 will be bound to a gravitating body. In General Relativity trajectories in the field of a black hole with energy E > 0 can approach on an unbound trajectory; if they don't get closer than $10 r_s$ they will remain unbound. However, for a non-rotating black hole between $3 r_s > r > 6 r_s$ there are unstable orbits that can loop the black hole once or several times. The exit direction will depend on the approach impact parameter, energy, angular momentum of the particle. (The whole subject of trajectories about a Schwarzschild black hole is somewhat involved, we shall not delve into here, see the excellent exposition in Chandrasekhar [11] chapter 3, and even more complicated for Kerr black holes [11].

Suppose that a K2 civilization can send a relativistic starship (slower than light, yet with a high Lorentz factor) in only a certain direction, because of the interstellar medium or some pointing advantage in a beamed energy system. If this K2 civilization has black holes mapped in the galaxy then a relativistic ship can be turned in the direction of the target by using this capture-unbound orbit mechanism with only a small expenditure of energy. It would demand that there is a K2 level of guidance, navigation and control and computational power to hit the right impact parameter. A vehicle can graze the distance of $3 r_s$ making many revolutions before exiting, but one must stay outside of $3 r_s$ or otherwise plunge into the hole. [10a, 11]. Setting u= 1/r for the Schwarzschild metric an ultra-relativistic particle, with impact parameter b, equations of motion can be written as [11]:

$$\left(\frac{du}{d\varphi}\right)^2 = 2Mu^3 - u^2 + \frac{1}{b^2} \quad [4]$$

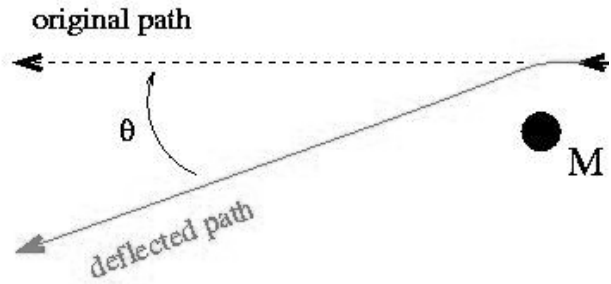

Figure 5. A relativistic particle deflected by a black hole.

Equation (2) has an approximate solution, if $b_c = 3\sqrt{3}M$, is the critical impact parameter and a particle approaches close to $b_c$ then the angle θ will become 'winding', that is it can orbit 0 to 2nπ times $\Omega$, Chandrasekhar [11].

$$b = 3\sqrt{3}M + 3.48e^{-\Theta} \quad [5]$$

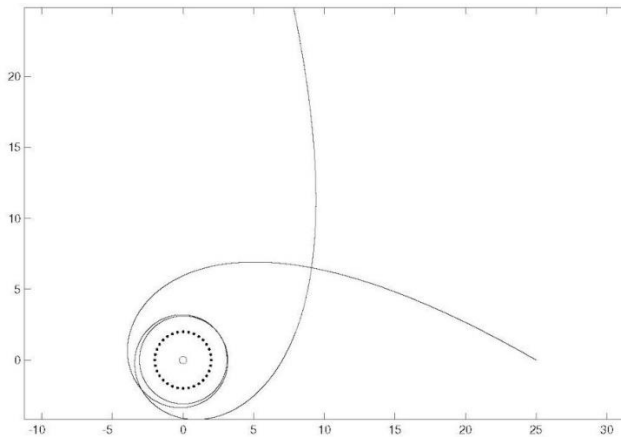

Figure 6. Winding orbit about a non-rotating black hole.

Of what advantage is this? First a K2 civilization might use such a capture orbit as a free source of direction change. A ship moving at, say, .5 c , would have to expend a lot of energy to change direction if a desired destination is not along a given trajectory. That is some fraction of the ship's total energy $E = \gamma mc^2$ would be needed to turn it. For instance, for a 1000 metric ton starship, E ~ $10^{13}$ terawatts, thus some fraction of that will be needed to turn it. A ~ 3 solar mass black hole can turn it for free. Why not move in that direction in the first place? That might be possible, but a ship may be constrained to a 'take off' path not in the target direction. Alas, if the black hole is in the vicinity of a target destination it would not be possible to use the fact that an orbiting particle close in a black hole will lose energy to gravitational radiation. Energy loss by gravitational radiation goes like $\Delta E = \frac{m^2}{M} f$ per orbit where f ~ 1, the mass of the ship, m, will much smaller than the mass of the black hole, the ship would have make ~ $10^{19}$ orbits!

To use this mechanism would require K2 technology capable of calculating the right impact parameter and have the shielding to survive the close by environment which may be an accretion disk (though there should be some 'bare' black holes in the universe). Kerr black holes will be the most common present extreme astrophysical environments (note: almost all stars that collapse to black holes will be rotating). For Kerr black holes such orbits exist but analytic calculations are extremely difficult [11] and will most likely have to be made numerically. Any K2 civilization 'hot' starship orbiting a Schwarzschild or Kerr black hole will have its waste radiation focused. Thus whenever an observer is in the line of sight a close orbiting object will have a fluctuating emission, peaked in the observer's direction. A starship looping a black hole like this would have an odd observational characteristic.

### III Bow Shocks

The use of magnetic fields for interstellar flight, first considered as a 'scoop' by Bussard [13]. Sagan suggested magnetic scoops this was extended to Mag-Sails by Andrews and Zubrin [14] who consider using them as 'brakes'. A magnetic field plowing into the interstellar medium (particularly dense regions) will incur both energy and momentum loss, noted by Bussard, quantified by Fisback in 1969[15]. This can be useful in stopping or at least slowing down a relativistic interstellar spacecraft. The byproduct of this process can produce a Bow Shock. Runaway neutron stars show such a structure,

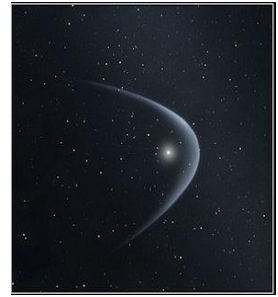

Figure 7: Neutron star bow shock analog for a starship. (source Wikipedia)

Radiation from the bow shock can range from the optical to the x-ray bands mostly produced by synchrotron radiation. A starship will be much smaller than a neutron star thus flux smaller, but it observation could imply a very peculiar object.

## IV Black Hole Lensing

If K2 civilizations utilize black holes as a method of redirection or as 'brakes' using gravitational radiation by orbiting in the non-Newtonian zone then the waste heat of the ship will be focused by the black hole one should see an anomalous peak in whatever part of the spectrum emerges from the black body radiation. A word of caution, strong gravitational field focusing is very complicated, where by 'strong' we mean the use of a Schwarzschild or Kerr black hole to bend light as a gravitational lens.

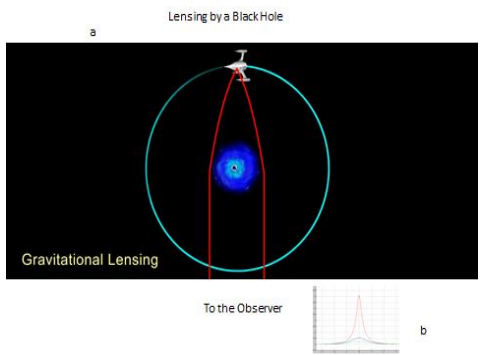

Figure 8 . Lensing of a starship's waste radiation by a black hole. (a) is the lensing, (b) image in the optical plane. (Douglas Potter)

### IV.1 Black Hole Beamed Propulsion

Consider a K2 civilization taking advantage of a Schwarzschild or Kerr black hole as a means of focusing radiation from a beaming station onto a sail. The advantage of this is the enormous amount of amplification possible. One of the most promising modes of interstellar flight propulsion methods is the use of a sail , a transmitter and maybe a 'lens' to focus a beam of laser light or microwaves. Extrapolate to a K2 civilization using a black hole as the focusing device. An approximate calculation for a Schwarzschild black hole shows that beamed radiation can be amplified by a factor $10^5$ to $10^{15}$. Caution is now advised. Almost all of the many astrophysical papers on 'strong focusing' consider a lens that is either a Schwarzschild or Kerr black hole, but in that case the source is either many light years away or is in orbit about the black hole but is physically larger in extent than the black hole. These constraints, though a realistic astronomical configuration, may not match the K2 technological engineering set up considered here. There are physical consequences to consider too. A source behind a Schwarzschild black hole does not come to focus at a point but creates , in the first approximation , on the optical axis (the axis that connects the source and the observers) , a 'caustic' where the amplification is extreme [16], [17].  A caustic, in the Schwarzschild case may be thought of as a 'tube' on the optical axis. This is because of the non-Newtonian nature of the strong gravitational field of a black hole. Photons that come from the right direction can go into orbit either permanently or for a finite number of revolutions as described above. With focusing the location of the source image will be displaced on the image plane. In Weak Lensing there will be an Einstein ring that is the deformation of the light from a source into a ring through gravitational lensing of the source's light by an object with an extremely large mass; black holes are the lenses of interest here. In the case of Kerr black holes the 'caustics' will be 'sheets' complicating the process to the extreme.

The exact location of a source and the sail location are the subject of further study, Figure 9.

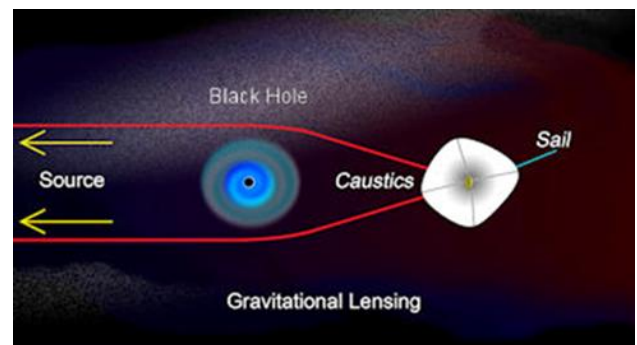

Figure 9.  Black hole gravitational lensing as beamed propulsion. (Douglas Potter)

## V . Zero rest or near zero rest mass carriers

Observational SETI has concentrated on using electromagnetism as the carrier, namely radio waves and laser radiation.  Michael Hippke [18] has pointed out that it may be possible to use neutrinos or gravitational waves as signals. Gravitational waves demand the command of the generation of very large amounts of energy. Neutrinos, like gravitational waves, have the advantage of extremely low extinction in the interstellar medium. To make use of neutrinos an advanced civilization could use a gravitational lens as an amplifier. The lens can be a neutron star or a black hole.  As outlined above using wave optics one can calculate the advantage of gravitational lensing for amplification of a beam and along the *focal axis* it is exceptionally large. Even though the amplification is very large the diameter of the beam is quite small, less that a centimeter. This implies that a large constellation of neutrino transmitters would have to enclose the local neutron star or black hole to make an approximate isotropic radiator. The operational energy needed is about .01 Solar, this means that such a beacon would have to be built by a Kardashev Type II civilization.

Table 2 Zero Rest Mass or Near Zero Rest Mass Carriers

| Carrier | Rest Mass | Lifetime | Extinction | Sources |
|---|---|---|---|---|
| Photon | 0 | Stable | .001 | Beacons -Waste Heat -Star Ships |
| Neutrino | ~.001 | Oscillations Stable | ~0 | Beacons Beams |
| Graviton | 0 | Stable | 0 | Beacons |

V.2 Black or Neutron Star Hole Beacon

For a compact gravitating body the gravitational gain by lensing is proportional to the ratio of the Schwarzschild radius and transmitter wave length, $r_s/\lambda$ and it is shown [6,8] that for amplification $\lambda < r_s$. Suppose a K2 civilization deploys a laser transmitter in orbit about a black hole, this transmitter-black-hole-lens-amplifier comprises a beacon (or it could be a neutron star as the lens). Townes [12] has shown that at short wavelengths infrared is favorable for transmission at signals over interstellar distances. The exact mass distribution of black holes is unknown, but an estimate of stellar mass black holes from observations and stellar evolution, the mass, m, is in the range of 3 to 20 solar masses [9, 10, and 11], take 10 solar masses as representative. For a basic example take the signal to be transmitted at 1 micron, the near-infrared. Take the K2 civilization as having placed this transmitter about a black hole lens of mass of 10 solar masses then the gravitational lens gain is $1.2 \times 10^{11}$, Jackson 19]

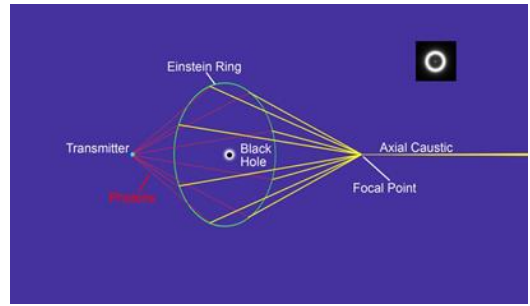

Figure 10. A schematic representation of gravitational lensing by a black hole. (Douglas Potter)

A one watt transmitter can reach a range of about 1 kpc (~ 3000 ly) and be detected within the magnitude 30 limit of the JWST. A laser transmitter alone would take an instrument with the sensitivity magnitude greater than magnitude 50 to detect.

V.3 Neutrino Beacon

To make use of neutrinos an advanced civilization could use a gravitational lens as an amplifier. The lens can be a neutron star or a black hole. Using wave optics one can calculate the advantage of gravitational lensing for amplification of a beam and along the *focal axis* and it is exceptionally large. Even though the amplification is very large the diameter of the beam, at the receiver, is quite small, less that a centimeter. This implies that a large constellation of neutrino transmitters would have to enclose the local neutron star or black hole to make an approximate isotropic radiator.

The engineering physics would be to build a constellation of neutrino beam transmitters. Place, in orbit, at 100 neutron star radii, $10^{18}$ advanced small Wakefield accelerators one meter long and 20 centimeters in diameter, figure 11. Then each point on figure 12 is occupied by an accelerator neutrino source, figure 11. Plasma-based accelerators are already producing high energy particle beams, what a K2 civilization may be capable of, for accelerators, is an extrapolation. With $10^{18}$ accelerators pointing four pi radians the probability of detection increases to approximately $10^{-3}$ at and Earth detector and the detection rate at 10,000 light years becomes approximately 5 per minute. The power required for the whole artifact' is about .01 Solar, which is a K2 command of energy. The operational energy needed is about .01 Solar; this means that such a beacon would have to be built by a Kardashev Type II civilization, Jackson, [20].

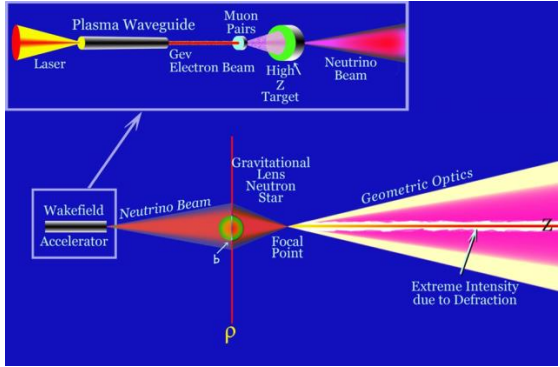

Figure 11. A gravitationally focused neutrino transmitter (Douglas Potter)

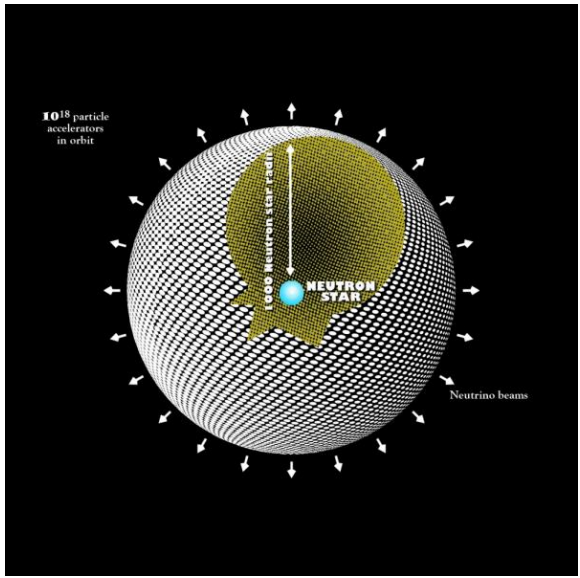

Figure 12. A schematic Constellation of $10^{18}$ neutrino accelerator-transmitters in orbit (nothing to scale). (Douglas Potter)

### V.4 Gravitational Wave Beacon

An advanced civilization might build a radiator to send gravitational waves signals by using small black holes. Micro black holes on the scale of centimeters but with masses of asteroids to planets might be manipulated by a super advanced instrumentality, possibly with very large electromagnetic fields. The machine envisioned emits gravitational waves in the GHz frequency range. If the source to receiver distance is a characteristic length in the galaxy, up to 10000 light years, the masses involved are at least planetary in magnitude. radiance. Back ground gravitational radiation sets a limit on the dimensionless amplitude that can be me measured at interstellar distance using an advanced LIGO like detector.

### VI. Gravitational Wave Transmitters

One could suppose that a civilization sends signals using gravitational waves. The LIGO receivers have seen gravitational radiation from natural objects. As a gravitational wave passes through matter it can change its geometry, namely its characteristic length. If one measures a length L and it responds to a gravitational wave by an amount ΔL, the 'strain' is measured by h= ΔL/L. This dimensionless amplitude is very small indeed, due to the weakness of gravitational waves. LIGO can measure h to the value of $10^{-22}$, or in approximate physical terms 1/1000 the diameter of a proton.

Table 3 : Advanced civilization gravitational wave transmitter located at 10,000 light years energy budgets.

| Dimensionless Amplitude h | Mass converted to Energy (ergs) | Kardashev Scale Civilization | Gravitational Wave Receiver |
|---|---|---|---|
| $10^{-22}$ | ~0.1 Earth Mass $10^{27}$ grams | 3.6 | LIGO at 100 Hz |
| $10^{-25}$ | ~ mass of Ganymede ~$10^{26}$ grams | 3.0 | Advanced Gravitational Wave Detector ~1GHz |
| $10^{-33}$ | ~ The mass of asteroid Ida ~ $10^{17}$ grams | 2.4 | 'Planck' Length Detector |

LIGO can detect a Type 3 plus civilization 100 light years away, but presently only in the frequency range of ~100 Hz. A more plausible signal, we argue, may lie in the GHz range. (In the following it is taken that a Kardashev civilization of a certain order means more than a mastery of a level of energy , that itself, implies an ability to project an instrumentality , engineering physics of staggering sophistication.)

Physically, h is related to the transmitter by h~ΔE/r where ΔE is a burst of gravitational radiation energy and r is the distance from the transmitter. Take ΔE as the amount of energy produced by the annihilation of a mass m, namely $mc^2$, and take the distance of the transmitter to be at 10000 light years (approximately the scale of the galaxy). The amount of energy produced can be related to the quantity of specified by the Kardashev scale.

To configure a GW machine, suppose an advanced civilization has planetary size black holes in its inventory. Four (or more) of the small black holes become 'orbital machine', a large central mass plus an exciter mass is one component. One active element of the machine, the central and exciter black holes, form binary systems orbiting the home star. (All the 'small' black holes may be rotating, Kerr, types). See Figure 13.

To provide the energy for this system one posits a very advanced civilization that has also Kerr black hole as a compoent to provide a super-radiance power station, as proposed by Jackson and Benford [21], see this paper for details and section VII. Other gravitational wave beacons have been proposed Abramowicz, et. al, [22], and Rana Adhikari [23]

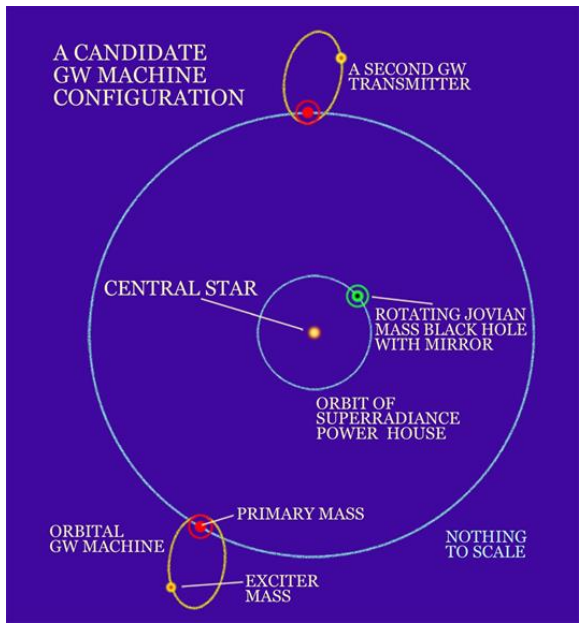

Figure 13. A schematic representation of a total 'gravitational wave machine' system. Note the 'Superradiance machine' is inward of the gravitational wave machines; the Home Star is located somewhere in this system.

VII. Black Hole Bomb Beacon

An electromagnetic wave impinging on a Kerr black hole can be amplified as it scatters off the hole if certain conditions are satisfied giving rise to and amplified wave called superradiant scattering [24]. By placing a mirror around the black hole one can make the system into a bomb! In the modeling a wave with frequency $\omega < m\Omega$ impinging on Kerr black hole will be amplified ( m is an azimuthal wave number and $\Omega$ the angular velocity of the Kerr hole at the horizon) (The azimuthal number is a number for the wave that determines its orbital angular momentum.) The scattered electromagnetic wave will be amplified, the excess energy being drawn from the Kerr hole's rotational energy.

If a K2 civilization builds a 'mirror' about a Kerr black hole undergoing this process the radiation will be amplified exponentially until the mirror fails are the radiation is released. The mirror cannot be a solid shell since that would be mechanically unstable. It would be an orbiting ensemble similar to a Dyson swarm. The orbits could be an oscillating shell the technology keeping it in configuration at a K2 level.

Consider a mirror assembled from a large number of elements of a truncated icosahedron, figure 14, it might be some other solid as long as the inside surface forms a mirror. As long as the configuration is such that transmitter reflectors located towards the Kerr black hole can efficiently contain the scattered radiation. The process would be that the transmitters fire once and then by K2 technology become reflectors, then the initial radiation would be amplified until the strength of the K2 'mirror-ships' artifact can no longer contain the electromagnetic energy and release it through ports.

Consider a 1 solar mass black hole rotating at about 10,000 radians per second, one can calculate the critical distance for spherical 'super radiant' mirror [24a]. It is located at 22 km (the event horizon is approximately at 3 km). When in operation at the end of 13 seconds the energy content is $10^{17}$ times the initial pulse. To match the 'bomb' constraints the transmitted pulse wavelength should be at about 18 km. How one would reflect and absorb long wave length radio waves is a problem to be solved by a K2 civilization. A possibility is that a spectrum of primordial black holes (PBH) exist left over from the Big Bang. PBH's in the range of $10^{-5}$ to $10^{43}$ grams might exist. For an Earth mass Kerr black black hole with event horizon 9.0 mm, placing the mirror at 1m one gets a growing timescale of about 0.02 seconds the critical radiation would be high frequency radio wave at about 33 GHz.

With amplification factors of the order $10^{17}$ one has K2 civilization solving the containment mirror problems, keeping system from melting or being shattered. It would mean the system would have to be fine-tuned to these effects. With the right configuration the structure would hold the energy until some material strength is exceeded while keeping the radiation absorption from vaporizing and the fleet of transmitters fly off with some fraction of the amplified energy. One watt in $10^{17}$ watts out! As a beacon it would be an unusual object.

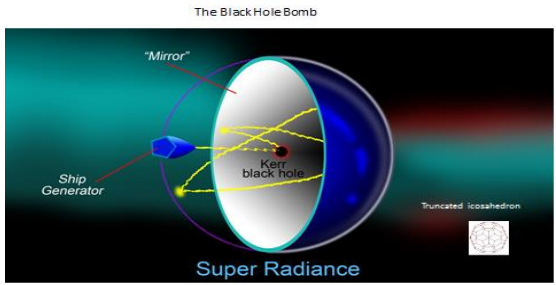

Figure 14. Black Hole Bomb (Douglas Potter)

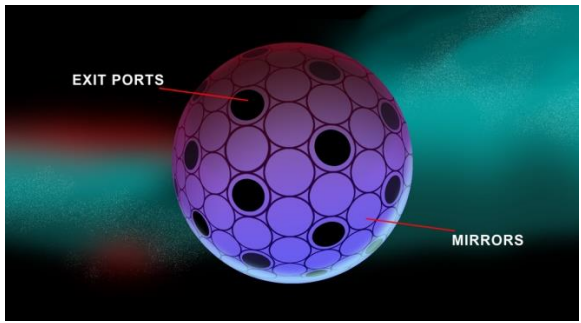

Figure 15. Black Bomb as a beacon. The beacon configuration is in figure XII , exit ports allow beams of amplified radiation out as signals. (Douglas Potter)

VIII. Megastructures

In 1960, Dyson Freeman described how the exponential rise in energy requirements by a technological civilization might lead to the construction of a Dyson Sphere around a star [4a]. This is a hypothetical mega-structure encapsulating a star in order to completely capture its energy output. A habitable surface would offer the additional bonus of having extra space for a continually expanding civilization. Therefore, the discovery of such an object would be an indicator of intelligent life.

Given the great number of observatories that have surveyed the sky, it can be said, relatively safely, that with more stars measured more accurately than ever before, zero Dyson spheres have been found at the present time.

There may yet be intelligent aliens out there, building vast trans-planetary empires to collect and utilize as much energy as possible, but the evidence for them is nil thus far [1].

A galaxy filled with Dyson Sphere might appear as a Kardashev III , this has been looked for and not found [25].

An extreme artifact that has been envisioned is the Shkadov thruster [26].

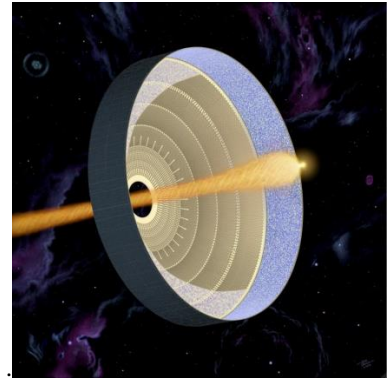

Figure 16. Shkadov thruster (Don Davis)

The concept being takes it planetary system on a galactic voyage the whole world of planets becoming a starship [26]. This would make a very unusual observational object.

Traversable wormholes might not be megastructures but they might be observable. Much has been written about the concept of traversable wormholes as 'faster than light' transport about the universe [27]. Traversable Wormholes would be extreme technological objects , possible K3 level, if constructed they might be observable by means gravitational lensing of light. 'Warp bubble' transport might also display observable effects [28], maybe gravitational shock waves or worse destruction of destination!

VIII Conclusions

We have presented some exotic techno-signatures attributed to advanced civilizations. At the moment SETI focuses almost solely on electromagnetic signatures with favor given to directed signals, beacons or possibly ambient leakage. This is likely the best way forward; however we have pointed to possible signatures that are not directed. Interstellar transport may have a detectable component. Megastructures other than Dyson spheres may have an observable existence also. Of interest is other carriers of information that may be directed such as neutrinos and gravitational waves. Since there are more instruments being built or planed for doing non-electromagnetic astronomy it is of interest to be mindful of possible anonymous signs received in that observable data.

References:

[1] NASA and the Search for Technosignatures: A Report from the NASA Technosignatures Workshop. https://arxiv.org/abs/1812.08681

[2] Cocconi, Giuseppe, and Philip Morrison. "Searching for interstellar communications". Nature 184 (4690). pp. 844–846. (1959).

[3] Freemann J. Dyson "Search for Artificial Stellar Sources of Infra-Red Radiation". Science 131 (3414): 1667–1668 (1960).

[4] Kardashev, Nikolai "Transmission of Information by Extraterrestrial Civilizations". Soviet Astronomy ,8: 217. (1964).

[4a]. Wright, Jason T., "Dyson Spheres", Serb. Astron. J. } 200 . 1-18, (2020)

[5] D. R. J. Viewing, C. Horswell, E. W. Palmer, "Detection of Starships," JBIS, 30, 99-104 (1977),

[5a] R. Zubrin "Detection of Extraterrestrial Civilizations via the Spectral Signature of Advanced Interstellar Spacecraft," Progress in the Search for Extraterrestrial Life, ASP Conference Series Vol. 74 (1995)

[6] Hale Bradt, Astrophysics Processes, Cambridge University Press (2008)

[7] Benford, James N. and Benford, Dominic J, Power beaming leakage radiation as a SETI observable, ApJ...825..101B, 2016.

[8] "Gravitational Machines." Interstellar Communication, A. G. W. Cameron, Editor, New York: Benjamin Press, 1963, Chapter 12.

[9] Pfahl, E., Podsiadlowski, P., and Rappaport, S., "Relativistic Binary Pulsars with Black Hole Companions",Astrophys. J.,628, 343–352, (2005)

[10] Y.B. Zeldovich and I.D. Novikov, Relativistic Astrophysics (Book 1), University of Chicago Press (September 30, 1971)

[11] Chandrasekhar, S. , The Mathematical Theory of Black Holes. Oxford: Clarendon Press.(1992).

[12] G. Benford, Bow Shock, 2007, http://www.baenebooks.com/chapters/1416521364/1416521364___4.htm

[13] Robert Bussard, "Galactic Matter and Interstellar Flight," Astronautica Acta Vol. 6 (1960)

[14] D. G. Andrews and R. Zubrin, "Magnetic Sails and Interstellar Travel", Paper IAF-88-553, 1988

[15] J. F. Fishback, Relativistic interstellar spaceflight, Acta Astronautica 01/1969

[16] Petters, Arlie O.; Levine, Harold; Wambsganss, Joachim. Singularity Theory and Gravitational Lensing. Progress in Mathematical Physics 21,(2001).

[17] J. A. H. Futterman, F. A. Handle and R. A. Matzner: Scattering from black holes. Cambridge University Press, Cambridge 1988.

[18] Hippke, Michael, Interstellar communication. IV. Benchmarking information carriers, https://arxiv.org/pdf/1711.07962.pdf

[19] A. A. Jackson, Black Hole Beacon: Gravitational Lensing, , JBIS, 68, pp.342-346,2015.

[20] A.A. Jackson, A Neutrino Beacon, JBIS, 73, pp.15-20, 2020.

[21] Jackson, A. A.; Benford, Gregory, A Gravitational Wave Transmitter, JBIS Vol 72 No 02 – February 2019

[22] Marek Abramowicz, Michal Bejger, Eric Gourgoulhon, Odele Straub, The Messenger: a galactic centre gravitational-wave beacon, arXiv:1903.10698,2019

[23] Rana Adhikari, Gravitational Wave Signatures, talk given at NASA Technosignatures Workshop, 2018.

[24] Press, William H.; Teukolsky, Saul A. "Floating Orbits, Superradiant Scattering and the Black-hole Bomb". Nature , 238(1972).

[24a] Vitor Cardoso, Óscar J. C. Dias, José P. S. Lemos, and Shijun Yoshida , Black-hole bomb and superradiant instabilities, Phys. Rev. D 70, 2004.

[25] Griffith, Roger L, et al., The Ĝ Infrared Search for Extraterrestrial Civilizations with Large Energy Supplies. III. The Reddest Extended Sources in WISE, The Astrophysical Journal Supplement Series, Volume 217, Issue 2, article id. 25, 34 pp. (2015).

[26] L.M. Shkadov, *Possibility of controlling solar system motion in the galaxy*, 38th Congress of IAF, October 10-17, 1987, Brighton, UK. Paper 1AA-87-613.

[27] Thorne, Kip, The Science of Interstellar, W. W. Norton & Company, 2014.

[28] John G. Cramer, Robert L. Forward, Michael S. Morris, Matt Visser, Gregory Benford, and Geoffrey A. Landis, Natural Wormholes as Gravitational Lenses,Phys. Rev. D 51, 3117 ,1995.